# Vanishing of Cyclotron Resonance in Correlated 2D Electron Systems

Andre Chebotarev and Galina Chebotareva

*PhysTech Lab*
*P.O. Box 20042 Stanford CA 94309 USA*
*Andrey@Chebotarev.net*

Experimental measurements of photoresistivity under terahertz (THz) radiation in low magnetic fields at conditions of cyclotron resonance (CR) in two-dimensional electron system (2DES) of GaAs/AlGaAs nanostructures are presented and discussed. We report the experimental discovery of "CR-vanishing effect" (CRV) in GaAs/AlGaAs heterostructures with high mobility as a well-defined gap on CR-line that is independent on incident THz power. Our analysis shows that the CRV may appear in systems with well correlated state of 2D electrons such as plasma waves and others. Fundamental nature of these correlated states of electrons in 2DES is discussed. Future THz detectors utilizing the new correlated states in 2DES may expand horizons for supersensitive detection in sub-THz and THz frequencies ranges.

*Keywords:* terahertz; cyclotron resonance; plasma wave; correlated electrons; 2DES; nanotechnology; THz detector; THz electronics; GaAs; graphene; Ge; Si; GaN.

## 1. Introduction

Future of THz electronics is closely connected with development of new nanomaterials based on well explored semiconductors: Si, Ge, GaAs, GaN and others. The emerging nanotechnologies as we believe will allow creating of new nanostructures from these wide-band-gap semiconductors to increase both working frequency and temperature of semiconductor devices at the same time. These nanotechnologies will permit to support and to grow up the semiconductor industry based on the modern wafer's processing technologies.

The more convenient way to demonstrate bright prospective of THz nanoelectronics is to create new THz detectors [1-5] and sources [8-10], because of their simplest design. The direction of new THz sources and detectors is being vigorously developed now [5]. We are working to make available a new supersensitive detector for sub-THz and THz frequencies ranges [1,2].

Modern THz detection technologies are based on the general idea of plasma waves as collective states of Two-Dimensional Electron System (2DES) in semiconductor nanostructures [1,2,7,11-15]. Our specific approach, based on our experimental study [1-6] of 2DES in low magnetic field in GaAs/AlGaAs heterostructures with the cyclotron resonance (CR) technique, allows to find [3,4] and separate single-electron-gas state (uncorrelated 2D electrons), from electron plasma waves and other correlated 2D electrons states.

In presented work we investigated nanostructures GaAs/AlGaAs, in which the single-electron-gas or uncorrelated 2D electrons effects in 2DES had been well studied by this time. The main focus in our work had being devoted to the structures with high electron mobility and correlated 2D electrons effects. These nanostructures are prospective for study and applications as future THz detectors: in these structures the electron - electron interaction may dominates at some low temperatures and that is why the effects caused by correlated 2D electrons will be brightly demonstrated and utilized for detection.

## 2. Experimental study of 2DES GaAs/AlGaAs structures in low magnetic field

We have measured cyclotron resonance (CR) and Shubnikov-de Haas (SdH) oscillations induced by THz radiation on samples of GaAs/Al$_x$Ga$_{1-x}$As structures (MBE), with electron density $3 \div 5 \times 10^{11}$ cm$^{-2}$, mobility $10^5 \div 10^6$ cm$^2$V$^{-1}$s$^{-1}$ at low magnetic fields up to 2 T, at temperature 4,2K$^0$ and frequencies $0.13 \div 0.15$ THz by photoresistance technique (Fig.1-3). Photoresistance ($\partial$R, a.u) is measured as magnetospectrum on magnetic fields (B, Tesla) under incident THz radiation (P$_{THz}$, $\omega_{THz}$), i.e. $\partial$R,a.u = F(B, P$_{THz}$, $\omega_{THz}$). Traditional *dc* measurements of Quantum Hall (QH) resistance ($R_{xy}$) and longitudinal resistance ($R_{xx}$) were also made and compared with differential *ac* measurements of cyclotron resonance and Shubnikov-de Haas (SdH) photoresistance ($\partial R$, a.u.) under THz radiation.

## 2.1. *Photoresistance magnetospectra under THz radiation*

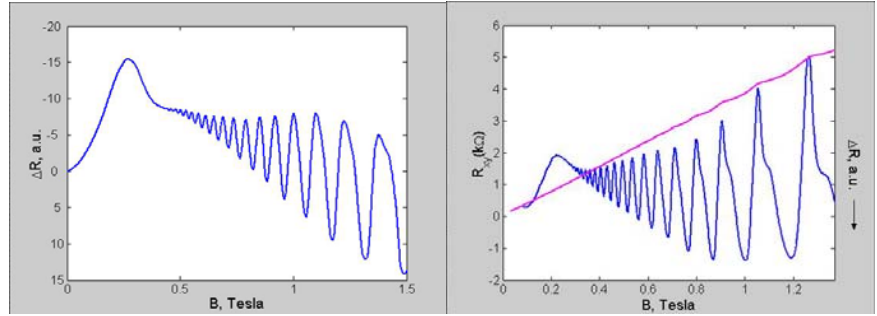

Fig.1 Photoresistance magnetospectrum measurement on 2DES in GaAs/AlGaAs nanostructure in conditions of cyclotron resonance and Shubnikov-de Haas oscillations under THz radiation.

Fig.2 Link up measurements of Quantum Hall resistance $R_{xy}$ (magenta curve) and THz photoresistance magnetospectrum in conditions of cyclotron resonance and Shubnikov-de Haas oscillations on the same GaAs/AlGaAs sample.

Fig.1 represents magnetospectrum of well-tinted line of CR-resonance and SdH photoresistance ($\partial R$, a.u) oscillations under THz radiation. Magnetospectrum on Fig. 2 demonstrates very sharp line of CR with regular SdH photoresistance oscillations that are resolved up to the maximum of CR-line. The CR-line maximum is shifted into the higher magnetic fields with increasing of THz radiation frequency in accordance with equation:

$$\omega_{CR} = eB/m^*c,$$

where **e** is electron charge, $m^* \approx 0,067\, m_e$ is effective mass of electrons in GaAs, $m_e$ is free electron mass, $\omega_{CR}$ is CR frequency.

At the same time the positions of the extreme points of SdH oscillations in magnetic field were found in our experiments as frequency independent.

Fig.2 compares Quantum Hall resistances and THz photoresistance magnetospectrum measurements in the conditions of cyclotron resonance and Shubnikov-de Haas oscillations on the same GaAs/Al$_x$Ga$_{1-x}$ As sample. The both methods provide complementary information about parameters of 2DES that allows conducting aim-directed research of new nanomaterials for high speed electronics.

In general the photoresistance magnetospectra presented on Fig.1 and Fig.2 are quite well explained by established concept about properties of 2 DES in GaAs/AlGaAs structures as the single-electron-gas state (uncorrelated 2D electrons).

The photoresistance magnetospectra were observed on samples with various electron concentrations and mobility, and other structure parameters. Obtained information can allow developing more optimal and target-defined growth of nanostructures for THz electronics.

To study correlated 2D-electrons effects we have conducted detailed measurements of nanostructures with maximal mobility and cooled down to cryogenic temperatures, when the scattering on photons and impurities becomes less and electron's momentum relaxation time increases enough to demonstrate effects in 2DES based on correlated electron states as plasma waves and others.

In some samples at specific conditions we have registered very complex photoresistance magnetospectra that reflecting complexity of 2DES. One from the most interesting effects has been observed by us is well-defined gap formation on CR-line named by Vanishing of Cyclotron Resonance (CRV).

## 2.2. "CR-vanishing effect" (CRV) in 2DES GaAs/AlGaAs

We report the experimental discovery of "CR-vanishing effect" (CRV) in GaAs/AlGaAs structures with high mobility in low magnetic field as a deep gap on CR-line that is independent on incident THz power (Fig 3).

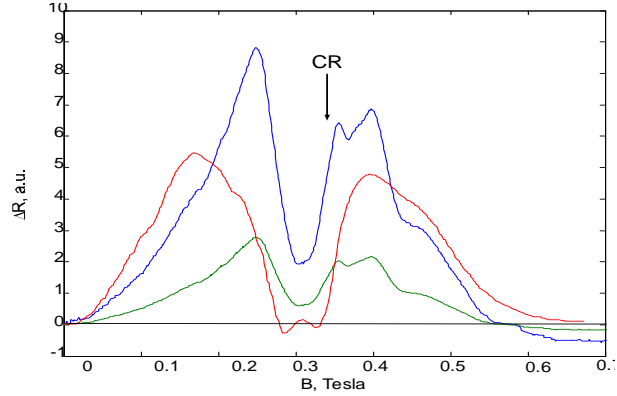

Fig3. "CR-vanishing effect" (CRV). The photoresistance magnetospectra of 2DES in the area of cyclotron resonance, where the photoresistance in the center of CR-line drops to minimum instead going to maximum (in comparison with Fig.1,2). Red curve vs. green & blue curves is measured at different THz frequency that is in ratio 0, 85 smaller; green curve is measured at reduced THz radiation power on 10 dB from blue curve.

It was observed that in wide enough area of low magnetic fields ($B_{gap}$) the photoresistance $\partial R(B_{gap})$ in the center of CR-line jump down to minimum instead going to maximum as in usual CR-lines and creates deep gap, where

$$\partial R(B_{gap}) \approx \partial R(B_0)$$

That looks like 2DES in gap magnetic fields ($B_{gap}$) has back to beginning state $\partial R(B_0)$. Thus, formally it is possible to suggest that 2DES in gap magnetic fields ($B_{gap}$) has become not affected by magnetic fields or THz photons, or momentum relaxation time become independent from electron energy. Other worlds, THz photon – 2D electron interaction has become very weak in $B_{gap}$ region. This is very unusual phenomenon in CR region for single-electron-gas (uncorrelated 2D electrons), where electron – THz photon interaction has maximum.

In these measurements (Fig.3) a sample was placed in Faraday configuration, i.e. perpendicular to magnetic field direction (blue & green curves) and was tilted about $30^0$ with magnetic field direction (red curve). Thus, for red curve the perpendicular component of total magnetic field is 0,85 B(Tesla), while for blue & green curves it is original B(Tesla). As was observed, CRV-effect is occurred in the same magnetic fields for red, blue & green curves, but on little bit shifted THz frequencies. THz frequency for red curve is different from THz frequency in measurement of blue & green curves in ratio 0,85 too. Thus, CRV-effect is proportional to the perpendicular component of magnetic field. This dependence on perpendicular component of magnetic field provides that CRV (such as CR) is determined by the orbital moments of 2D electrons, but not their spins that are depended on the total magnetic field.

To check and eliminate possible non-linear effects and sample heating, we reduced THz radiation power on 10 dB (green curve). This change does not affect on CRV effect!! The shape of magnetospectrum (green curve vs. blue curve) did not change, while amplitude $\partial R(B)$ is proportionally reduced. It means that incident THz radiation does not change 2DES properties, while examines them.

Measurements of Hall resistance $R_{xy}$ under action of non-modulated (CW regime) and modulated THz radiation were performed. It was found that Hall resistance there is in the CRV-gap area of magnetic fields and any changes of $R_{xy}$ were not observed. It means that 2DES in $B_{gap}$ has indeed subjected by magnetic fields and senses it. Measured Hall resistance is very close to typical $R_{xy}$ presented by magenta curve on Fig.2.

## 3. Discussion of the experimental data and theoretical approaches

### 3.1. *Relationship of Hall resistance & magnetospectrum of photoresistance*

Fig.4 demonstrates THz photoresistance magnetospectrum (copy blue curve from Fig.2) of cyclotron resonance and Shubnikov-de Haas oscillations measured on GaAs/Al$_x$Ga$_{1-x}$As sample and the derivative (green curve) from measured Quantum Hall resistance R$_{xy}$ presented by magenta curve on Fig.2. Fig.5 demonstrates the same as on Fig.4 QH-resistance derivative (green curve), but reversed on sing THz photoresistance magnetospectrum of CR and SdH (blue curve). This QH-resistance derivative is really similar SdH oscillations, if taken with minus sing or in reversed case (Fig.5). Other differences can give additional information for consideration. Relations suggested from these data is

$$\frac{\partial \text{Rxy}}{\partial \text{B}} \approx -\alpha \times \frac{\partial \text{R}}{\partial \text{P}_{THz}} \quad ,$$

where *α* is some constant, P$_{THz}$ is THz power.

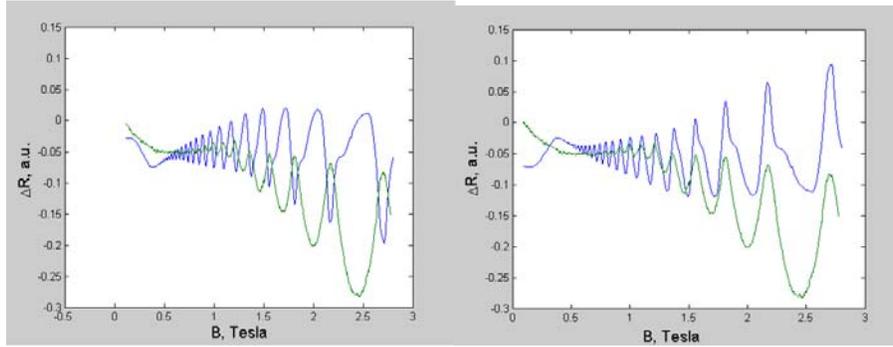

Fig. 4 and Fig.5  Analysis & correlation of Hall resistance & photoresistance magnetospectrum.

This comparison of THz photoresistance magnetospectra with derivative of Hall resistance allows considering many possible mechanisms of electron's scattering as phonons, impurities, electrons, and others. Our preliminary analysis shows that at cryogenic temperatures the electron-phonon interactions are not the main one and rather the Coulomb interactions are predominant.

### 3.2. *Vanishing of Cyclotron Resonance (CRV)*

Deep gap formation on CR-line named by Vanishing of Cyclotron Resonance (CRV) has been observed from our experimental study of 2DES GaAs/AlGaAs structures in low magnetic field by THz photoresistance technique. To check fundamental nature of the CRV effect we made detailed analysis of CRV-shape dependence on incident THz radiation power.

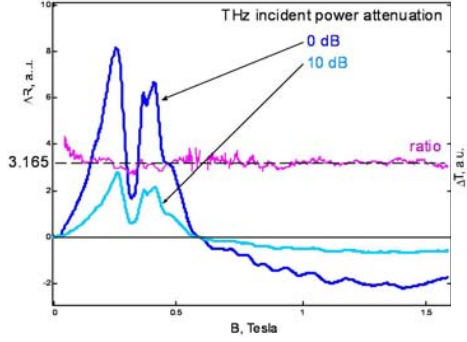 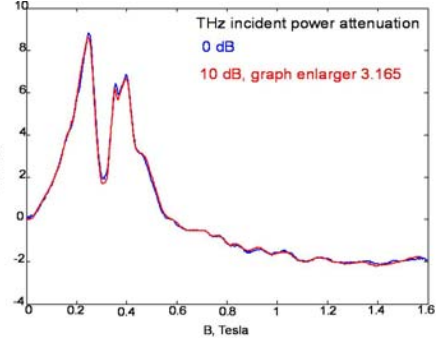

Fig.6 Measured CRV at different attenuations of incident THz power: 0 dB (blue) and 10 dB (cyan), and their ratio (pink).

Fig.7 Measured CRV without attenuations (0 dB) of incident THz power (blue) and calculated CRV curve by 3.165 times multiply of CRV-signal at 10 dB attenuation (red).

Fig. 6, and Fig.7 demonstrate that THz photoresistance magnetospectrum has become smaller in 3.165 times at 10 dB attenuation (3.165 is square root from 10): that makes obvious that THz photoresistance in low magnetic fields have linear dependence on amplitude, but not on power of incident THz electromagnetic waves. It is very important that shape of photoresistance magnetospectrum and CRV- gap itself do not depend on THz power or amplitude.

CRV-gap shape, where the photoresistance in the center of CR-line drops to minimum instead going to maximum is far from ideal CR-line shape. Lorenz shape of CR-absorption does not have any gap in approach of single-electron-gas so cannot provide a gap in photoresistance:

$$P(\omega) = \frac{\sigma_0}{2}|E|^2 \frac{1+(\omega^2+\omega_{cr}^2)\tau_m}{\left[1+(\omega^2-\omega_{cr}^2)\tau_m^2\right]^2 + 4\omega_{cr}^2\tau_m^2} \quad ,$$

where $\omega_{CR}$ is CR frequency, $\tau_m$ is electron momentum relaxation time.

Therefore, measured and analyzed data directs us to consider different physical models of behavior of two-dimensional electron system in nanostructures. While impossible to explain CRV by properties of single-electron-gas (uncorrelated 2D electrons) only, an appearance of some "correlated" states of 2D electrons has been suggested.

### 3.2. *Physical models of "correlated" states of 2D electrons in nanostructures*

Our experimental results demonstrate appearance of the fundamental correlated states of electrons in 2DES at low magnetic field. Theoretical analysis of the "CR-vanishing effect" indicates that a few physical models of the new fundamental correlated states of electrons in 2DES at low magnetic field may be discussed at this time. Here we consider magneto-plasma waves, a model for Quantum Hall effect, and non-linear zero-resistance states that have mentioned in literature on 2DES GaAs/Al$_x$Ga$_{1-x}$ As structures in the similar experimental conditions.

### 3.3.1. *Magneto-plasma waves*

Plasma frequency $\omega_p$ in our experimental conditions have been estimated from the equation

$$\omega_p^2 = 3\pi^2 n_{2D} e^2 /(2m^* \varepsilon_{eff} d) \quad ,$$

where $n_{2D}$– 2D electron density and $d$ – sample size. We got that $\omega_p \approx 1.5 \times 10^{-2}\, THz$.

The estimated plasma frequency $\omega_p \geq 0.1\, \omega_{THz}$, where $\omega_{THz}$ is frequencies 0.13÷0.15 THz used in our experiments. Nevertheless magneto-plasma resonance in area of CR may occur[7] and here is determined[16] by

$$\omega_{MPR} = \omega_{CR}/2 + [\omega_p^2 + (\omega_{CR}/2)^2]^{1/2}.$$

Then, magneto-plasma waves in correlated 2D electron systems may come about frequencies $\omega_{MPR} \approx 1.1\, \omega_{CR}$ and accordingly in magnetic fields $B_{MPR} \approx 0.9 B_{CR}$ i.e. very close to the maximum of CR ($B_{CR}$). The measured position of CRV-gap in magnetic field $B_{gap}$ is very well corresponded to 0.9 $B_{CR}$, or $B_{gap} \approx B_{MPR}$; thus a nature of CRV-effect could be explained by magneto-plasma waves of correlated 2D electrons. However, it is not clearly why does the magneto-plasma effect compensate or decrease the CR effect? The both effects should rather be additive to each other, then be concurrent.

### 3.3.2. *Comparison with models for Quantum Hall effect*

The experiments[17] showed that CR is also vanishing but at high magnetic fields only. In this works of K. von Klitzing and others[17] was mentioned that "the CR absorption process is not visible in photoconductivity, if the CR resonance locates in the plateau region" because of longitudinal resistance $R_{xx}$ is going to zero. The plateau regions on Hall resistance $R_{xy}(B)$ occur in high magnetic fields as Quantum Hall effect.

Measured by us, CRV-effect also indicates that the longitudinal resistance $R_{xx}$ is going to zero, but in the low magnetic fields ($B_{gap}$). The measured Hall resistance $R_{xy}(B)$ is linear at these low magnetic fields ($B_{gap}$) and is not quantized *vs.* $R_{xy}(B)$ at higher magnetic fields, where steps ("the plateau") on $R_{xy}$-curve arise due to Quantum Hall effect (typical $R_{xy}(B)$ see on Fig.2).

Thus, CRV-effect measured at not-quantized Hall resistance can not be explained by the Quantum Hall effect, while the longitudinal resistance $R_{xx}$ is going to zero at low magnetic fields too.

### 3.3.3. *Comparison with non-linear zero-resistance states*

The CRV-effect was observed in the comparable experimental conditions with work[18], where "microwaves induced vanishing resistance effect" was demonstrated and zero-resistance state has been suggested for 2DES GaAs/AlGaAs structures. This work shows that Hall resistance $R_{xy}(B)$ is not quantized at the low magnetic fields corresponding CR, i.e. similar to our experiments. Also the vanishing longitudinal resistance $R_{xx}$ at low magnetic field was demonstrated[18], while another measurement technique was used.

However, occurring of the "microwaves induced vanishing resistance effect" is strongly depended on incident THz power in contrast with our CR-vanishing effect (3.1.).
Moreover, in work[18] CR-line is not presented, while cyclotron resonance must be exposed by free electrons under action uniform electromagnetic and magnetic fields. As well, SdH oscillations that correspond to Landau quantization are disappeared with increasing of microwave power[18]. Finally, we have applied about the same THz power and frequencies, magnetic fields and temperatures on similar structures as in work[18] and could measure the both CR-line and SdH oscillations, thus we have found and study a different phenomenon.

### 3.3.4. *Other physics is considering*

Thus, at this moment we exclude models for Quantum Hall effect[17] and non-linear zero-resistance state[18]. Our specific approach involves separation of single-electron-gas state (uncorrelated 2D electrons) from electron plasma waves and other correlated 2D electrons states. Searching of current physical models of correlated states in 2DES or creating new one can be suggested at this moment to give explanation to the CR-vanishing effect and others from ours study of 2DES in GaAs/AlGaAs structures by CR-THz photoresistance and transmission techniques.

## 4. Future THz detectors

Our study of 2D electron systems in GaAs/AlGaAs structures, having fundamental physics interest, is directed to develop new kind of THz devices. The research was made on nanostructures with various electron concentrations and mobility, and other structure parameters. Obtained information allows conducting aim-directed research of new nanomaterials for high speed electronics and developing more optimal and target-defined growth of nanostructures for THz electronics.

Appearance of correlated states in 2DES may be considered as next step for supersensitive detection in sub-THz and THz frequencies ranges. Correlated states may form of sub-bands structure and subsequently arise of resonant electron transitions between sub-bands under action of electromagnetic quanta. Crosssection of absorption for resonant electron transitions larger then crosssection of absorption for nonresonant electron transitions on a few orders in THz frequencies range. Thus appearance of correlated states in 2DES may considerably increase sensitivity of sub-THz and THz detectors.

Also, electron's transition from correlated state to uncorrelated state may significantly change sample's resistance. Wide known example is superconductor detectors. Comparison with well developed superconductor detectors (that exploits correlated electron's states) predicts that NEP value may be below $10^{-15}$ W*Hz$^{-1/2}$ at higher working temperature in semiconductor nanostructures due to utilizing correlated electron's states of 2DES[1].

Unique experimental approach allows us investigate property of 2DES in the central area of sample without boundary effects, for example edge magnetoplasmons. Recently investigation of edge magnetoplasmons propagating along a boundary of 2D charged system and based on this effect tunable detector and solid state spectrometer was demonstrate[19]. Actually, propagating of edge magnetoplasmons is rather one dimensional (1D) effect. Therefore, our detection technology has principal advantage in sensitivity and selectivity because density of 2D electron states much larger then 1D one. Additionally, interaction of electromagnetic quanta with whole sample's area more effective then interaction with sample's boundary.

Future of THz electronics is closely connected with development of new nanomaterials based on well explored semiconductors: Si, Ge, GaAs, GaN and others. The emerging nanotechnologies as we believe will allow creating of new nanostructures from these wide-band-gap semiconductors to increase both a working frequency and a temperature of semiconductor devices at the same time. Thus, this kind of materials promises maximal advantages and effectiveness among other approaches to materials for THz electronics[20]. These nanotechnologies will permit to support and to grow up the semiconductor industry based on the modern wafer's processing technologies also.

## 5. References

1. A. Chebotarev & G. Chebotareva, THz Detection by Correlated 2D Electron Systems in *Proceedings of IRMMW-THz 2008*, Pasadena, California, USA, 15-19 Sept. 2008. ISBN:978-1-4244-2119-0
2. A. Chebotarev & G. Chebotareva, Cyclotron Resonance Vanishing Effect and THz Detection, *Proceedings of WOFE-07* Mexico, Dec. 2007. Intern. Journal of High Speed Electronics & Systems (2008) in press
3. A.P. Chebotarev & G. P. Chebotareva, Terahertz Induced Photoconductivity of 2D Electron System in HEMT at Low Magnetic Field in *Proceedings of ICPS-27*, Flagstaff, Arizona, 2004, AIP, 772, 1184-1185 (2004).
4. A. Chebotarev & G. Chebotareva, http://arxiv.org/ftp/cond-mat/papers/0411/0411356.pdf.
5. A. P. Chebotarev & G. P. Chebotareva, Commercialization of THz-IR photonics for homeland security and material inspection *in CLEO/PhAST Conference*, San Francisco, 2004.
6. A.P. Chebotarev, G.P. Chebotareva, A.P. Nikitin & E. Gornik, Deformation of cyclotron resonance spectrum of 2D-electron gas in GaAs/AlGaAs heterostructures. in *Inter. Symposium "Nanostructures : Physics and Technology"*, Sankt-Petersburg, Russia, 38-41 (1995).
7. V. I. Gavrilenko, V.N. Murzin, S.A. Saunin & A.P. Chebotarev, Magnetoplasma resonance in electron-hole drops in Germanium at arbitrary crystallographic orientation of the magnetic field, *Sov. Phys. Lebedev Inst. Report*, **3**, 11-17 (1978).

8. A. P. Chebotarev & V. N. Murzin, Millimeter-range emission from hot electrons in n-type Ge in crossed electric and magnetic fields, *Sov. Phys. JETP Letters*, 40, 1005-1008 (1984).
9. A. P. Chebotarev & V. N. Murzin, Generation and spectral composition of stimulated longwave infrared radiation of hot holes in Germanium in strong E_1_H fields, *Sov. Phys. Lebedev Inst. Report*, **5**, 27-31(1986).
10. Yu. A. Mityagin, V. N. Murzin, S.A. Stoklitskiy, A. P. Chebotarev & I.M. Melnichuk, Wide-range tunable submillimetre cyclotron resonance laser, *Optical and Quantum Electronics*, 23, 307-311(1991).
11. M. I. Dyakonov & M. S. Shur, Detection, Mixing, and Frequency Multiplication of Terahertz Radiation by Two Dimensional Electronic Fluid, *IEEE Transactions on Electron Devices*, 43, 3, 380-387(1996).
12. V.Ryzhii, I.Khmyrova, M.Ryzhii, A.Satou, T.Otsuji, V.Mitin, & M.S.Shur, Terahertz plasma waves in two-dimensional electron systems and their device applications, *Int. J. High Speed Electron. Systems* (2007), in press.
13. V.Ryzhii, I.Khmyrova & M. Shur, Terahertz photomixing in quantum well structures using resonant excitation of plasma oscillations, *J. Appl. Phys.*, **91**, 4, 1875-1881 (2002).
14. V.Ryzhii, I.Khmyrova, A. Satou, P.O.Vaccaro, T. Aida & M. Shur, Plasma mechanism of terahertz photomixing in high-electron mobility transistor under interband photoexcitation, *J. Appl. Phys.*, 92, 5756-5760 (2002).
15. V. M. Muravev, C. Jiang, I. V. Kukushkin, J. H. Smet, V. Umansky & K. Von Klitzing, Spectra of magnetoplasma excitations in back-gate Hall bar structures, *Phys. Rev. B* **75**, 193307 (2007).
16. I.V. Kukushkin, J.H. Smet, K. Von Klitzing & W. Wegscheler, Cyclotron resonance of composite fermions, *NATURE*, 415, 24, 409-412 (2002).
17. D. Stein, G. Ebert, K. von Klitzing & G. Weimann, Photoconductivity on GaAs/AlGaAs heterostructures, *Surface Science*, 142, 406-411 (1984).
18. R. Fitzgerald, Microwaves Induce Vanishing Resistance in Two-Dimensional Electron Systems, Physics Today, **56**, 4, 24-27 (2003).
19. K. Von Klitzing, S. A. Mikhailov, J. H. Smet & I. V. Kukushkin, Detector for electromagnetic radiation and a method of detecting electromagnetic radiation, U. S. Patent 6987484, Jan. 2006
20. On line: www.thz.us
21. Andre Chebotarev & Galina Chebotareva, Vanishing of Cyclotron Resonance In Correlated 2D Electron Systems in *Abstracts of ICPS-29 29th International Conference on the Physics of Semiconductor*,2008, Rio de Janeiro, Brazil